\documentclass[twocolumn,showpacs,preprintnumbers,amsmath,amssymb]{revtex4}

\usepackage[cp1251]{inputenc}
\usepackage{graphicx}
\usepackage{dcolumn}

\begin{document}

\title{Electron-Positron Pair Production by an Electron \\
in a Magnetic Field Near the Process Threshold}

\author{Novak O. P.}
\email{novak-o-p@ukr.net}

\author{Kholodov R. I.}
\email{kholodovroman@yahoo.com}

\affiliation{Institute of Applied Physics, National Academy
of Sciences of Ukraine, Sumy, 40030 Ukraine}

\author{Fomin P. I.}
\email{pfomin@bitp.kiev.ua}
\affiliation{Institute of Applied Physics, National Academy
of Sciences of Ukraine, Sumy, Ukraine}
\affiliation{Bogolyubov Institute of Theoretical Physics,
National Academy of Sciences of Ukraine, Kiev, Ukraine}

\date{2010}

\begin{abstract}
The electron–positron pair production by an electron in a strong magnetic
field near the process threshold is considered. The process is shown to be
more probable if the spin of the initial electron is oriented
along the field. In this case, the probability of the process is
$\sim10^{13}\,\, s^{-1}$ when the magnetic field strength is
$H=4\cdot 10^{12}$~G.
\end{abstract}

\maketitle

\section{INTRODUCTION}
Investigating quantum-electrodynamic processes
remains topical in connection with the existence of
neutron stars with magnetic fields comparable to or
greater than the critical Schwinger field, $H_c\approx4.41\cdot10^{13}$~G
\cite{Shapiro}.

The production of electron–positron pairs is an important element in pulsar
models, because the presence of an electron-positron plasma is believed to be a
necessary condition for the generation of coherent radio emission.
Many theoretical works are devoted to explaining the absence of radio
pulsars with long periods, which may be due to the termination of pair
production. For example, the mechanisms of plasma generation by one and
two-photon photoproduction were considered in \cite{Harding2}.
The pair production by an electron can compete with these
processes in strong fields.

Magnetic fields of sufficient strengths are so far
unattainable in laboratory conditions. The record
constant and pulsed magnetic fields are $10^{6}$~G
\cite{Maglab} and $3 \cdot 10^7$~G \cite{Sakharov},
respectively. However, the pair production
by an electron was experimentally observed in a strong
laser field in SLAC (USA) \cite{Burke1}. As the authors of
\cite{Burke1} point out, no consistent quantum-electrodynamic
(QED) theory of this process has been constructed.

Note also that QED processes take place during the
collisions of heavy ions. If the impact parameter is
$10^{-11}$~cm, then the magnetic fields in the region
between the ions can reach $10^{12}$~G. We suggest that
such processes were observed in Darmstadt, GSI
(Germany) \cite{Koenig}. At present, the new FAIR project is
being built in GSI one of whose objectives is to test the
QED theory in strong electromagnetic fields. In principle,
experiments on the observation of QED processes in the magnetic
field produced by heavy ions can be carried out within
the framework of FAIR.

The electron–positron pair production by an electron
in a magnetic field was first mentioned in \cite{Klepikov, FIAN}.
Nevertheless, no consistent QED calculation of the
probability was performed. The cross-channel of this
process is the electron scattering by the electron \cite{Graziani}.

The goal of this paper is to calculate the probability
of pair production by an electron near the process
threshold in the context of Furry’s picture. In this
case, the magnetic field strength is close to the critical
$H_c$, but it does not exceed its value, so that

\begin{equation}
\label{h}
h=H/H_c \ll 1.
\end{equation}
We will restrict our analysis only to the cases where the
final particles are at the ground Landau levels.

\section{KINEMATICS}
\begin{figure}[h]
\resizebox{\columnwidth}{!}
{\includegraphics{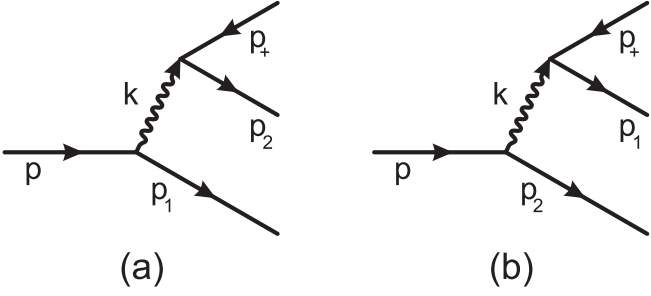} }
\caption{Direct (a) and exchange (b) Feynman diagrams for the
electron–positron pair production by an electron in a magnetic field.}
\label{Diagrams}
\end{figure}

The Feynman diagrams of the electron–positron
pair production by an electron are presented in Fig.~\ref{Diagrams}
The straight lines in the figure represent the solutions
of the Dirac equation in the presence of a classical uniform magnetic field.
In this case, the field strength is smaller in order of magnitude
than the critical one, $H_c\approx4.41\cdot10^{13}$~G.

Let us choose a coordinate system in which the
magnetic field  $\vec H$  is directed along the $z$  axis. The
eigenvalues of the electron energy in the magnetic
field are then
\begin{equation}
E_l= \sqrt{p_{z}^2+m^2+m^2 2lh}.
\end{equation}
Here, $l$ is the Landau level number and $p_{z}$ is the $z$
component of the electron momentum.

The magnetic field does not change in going to the
frame of reference that moves along the $z$ axis. Therefore,
without any loss of generality, the longitudinal
component of the initial electron momentum may be
set equal to zero: $p_z=0$. Consequently,
\begin{equation}
E=\tilde m= m \sqrt{1+2lh}
\end{equation}
for the initial electron.

The kinematics of the process is defined by the following
conservation laws:
\begin{equation}
\label{eq1}
\left\{
\begin{array}{l}
E_1+E_2+E_+=E,\\
p_{1z}+p_{2z}+p_{+z}=p_z=0,
\end{array}
\right.
\end{equation}
where  $E$ and  $p_z$ are the energy and longitudinal
momentum of the initial electron; $E_1$, $E_2$ and $E_+$ are
the energies of the final electrons and positron; $p_{1z}$, $p_{2z}$
and $p_{+z}$ are their longitudinal momenta.

First of all, note that the process is impossible if the
initial electron energy is insufficient for pair production.
It is easy to verify that the threshold condition is
\begin{equation}
\label{eq90}
\left\{
\begin{array}{l}
p_{1z}=p_{2z}=p_{+z}=0,\\
\tilde m =\tilde m_1 + \tilde m_2 + \tilde m_+.
\end{array}
\right.
\end{equation}

Generally, this condition cannot be met, because
the effective masses are discrete quantities. Thus, the
threshold values of the longitudinal momenta of the
final particles are nonzero. Expanding Eq.~(\ref{eq1}) into a
series of momenta, we will obtain the relation
\begin{equation}
\label{eq100}
\frac {p_{1z}^2}{b_1^2}+\frac {p_{2z}^2}{b_2^2}+\frac {p_{+z}^2}{b_+^2}=1,
\end{equation}
where
\begin{gather*}
b_f^2=2\tilde m_f \Delta, \\
\Delta=\tilde m-\tilde m_1-\tilde m_2-\tilde m_+,
\end{gather*}
and the subscript $f$ numbers the final particles ($f=1,\,2,\,+$).

It is easy to verify that at the process threshold when
\begin{equation}
\label{lf0}
l_f=0,
\end{equation}
the following conditions are met:
\begin{equation}
\label{pf}
\begin{array}{l}
\tilde m_f = m,\\
\Delta \le mh,\\
p_{fz}\lesssim m\sqrt{h}.
\end{array}
\end{equation}

\begin{figure}[h]
\begin{center}
\resizebox{0.9\columnwidth}{!}{
\includegraphics{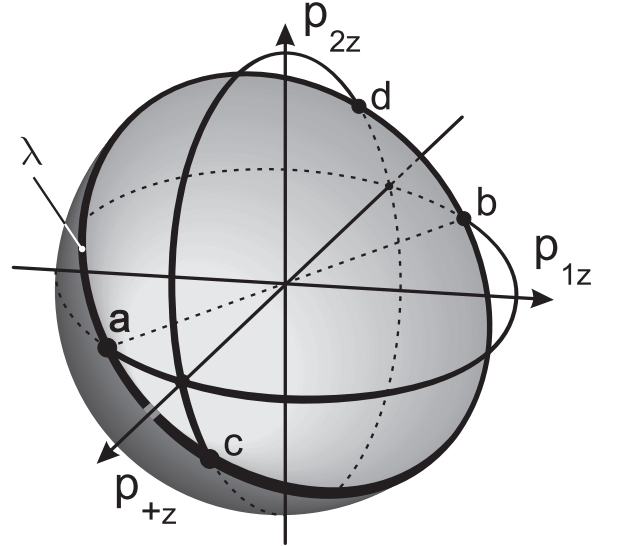}}
\caption{Threshold particle momenta -- the points of ellipse $\lambda$.
Points \textbf{a}, \textbf{b} and \textbf{c}, \textbf{d} are the points
of intersection of ellipse $\lambda$ with the $(p_{+z},p_{1z})$
and $(p_{2z},p_{+z})$ planes, respectively.}
\label{Kinematics}
\end{center}
\end{figure}

In the coordinate system where the momenta  $p_{1z}$, $p_{2z}$
and $p_{+z}$ are along the axes, the energy conservation
law (\ref{eq100}) specifies an ellipsoid. The possible values of the
momenta correspond to the points of the ellipse formed
by the intersection of ellipsoid (\ref{eq100}) with the plane
specified by the momentum conservation law (Fig.~\ref{Kinematics}):
\begin{equation}
\label{eq110}
p_{1z}+p_{2z}+p_{+z}=0.
\end{equation}

\section{THE PROBABILITY OF THE PROCESS}
According to general rules of quantum electrodynamics,
the probability amplitude for the process is
\begin{equation}
\label{general}
\begin{array}{l}
\displaystyle S_{fi}=i\alpha\int \!\! \int d^4xd^4x'\left[
(\bar\Psi_2 \gamma^\mu \Psi)D_{\mu\nu}
(\bar\Psi_1' \gamma^\nu \Psi_+') - \right.\\
\left.-(\bar\Psi_1 \gamma^\mu \Psi)D_{\mu\nu}
(\bar\Psi_2' \gamma^\nu \Psi_+') \right].
\end{array}
\end{equation}
Here, the prime on the wave function means that it
depends on the components of the primed 4-radius
vector $x'$. Let us substitute the wave functions \cite{Fomin2000} and
the photon propagator \cite{Landau4} into the amplitude:
\begin{equation}
\label{propagator}
D_{\mu \nu}=\frac{g_{\mu \nu}}{(2\pi)^4}
\int{d^4ke^{-ik(x-x')}\frac{4\pi}{k^jk_j}}.
\end{equation}
Since the dependencies of the wave functions on time
and $y$, $z$, $y'$, and $z'$ coordinates are the same in form as
those for plane waves, integration over these quantities
gives $\delta$-functions that express the energy and
momentum conservation laws. The integrals over the $x$ and $x'$
coordinates can be expressed in terms of special functions
studied in \cite{Klepikov,FIAN}. Substituting their explicit form
at $l_f=0$ yields the following expression for the
probability amplitude of the process:
\begin{equation}
\label{Sfi}
S_{fi}=S_1-S_2.
\end{equation}
Here,
\[
\begin{array}{l}
\displaystyle S_1=\frac{i\alpha \, 2 \pi^3 }{S^2 \sqrt{\tilde m E E_1 E_2 E_+}}
\frac{B_1^{\pm}\sqrt{\tilde m -\mu m}}{m\sqrt{l!h}}  e^{-a^2} X_1\times\\
\times \delta(E\!-\!E_1\!-\!E_2\!-\!E_+) \delta(p_{y}\!-\!p_{1y}\!-\!p_{2y}\!-\!p_{+y})\times\\
\delta(p_{z}-p_{1z}-p_{2z}-p_{+z}),
\end{array}
\]
$S_2$ is the exchange term,
\begin{equation}
\label{intX}
X_1=\int{\frac{(a+i\xi)^l}{\rho^2-\xi^2}e^{-\xi^2-2ib\xi}d\xi},
\end{equation}
\begin{equation}
\label{ab}
\begin{array}{l}
a=(p_y-p_{1y})/m\sqrt{2h},\\
b=(p_y-p_{2y})/m\sqrt{2h},\\
\xi=k_x/m\sqrt{2h},\\
\rho^2=\Omega^2-a^2, \quad \Omega^2=h/2,
\end{array}
\end{equation}
\begin{equation}
\label{bispinors}
B_1^+=4m\sqrt{m\tilde m}, \quad B_1^-=4p_{1z}\sqrt{m\tilde m}.
\end{equation}

We will obtain the probability of the process by
multiplying the square of the absolute value of the
amplitude by the number of final states:
\begin{equation}
\label{Wgeneral}
dW=|S_{fi}|^2\frac{Sd^2p_1}{(2\pi)^2}
\frac{Sd^2p_2}{(2\pi)^2}\frac{Sd^2p_+}{(2\pi)^2},
\end{equation}
where $d^2p_f=dp_{fy}\,dp_{fz}$.

Squaring the absolute value of (\ref{Sfi}) yields the
differential probability of the process per unit time
\begin{equation}
\label{Wtime}
\begin{array}{l}
dW=M\left|e^{-a^2}X_1B_1^\pm - e^{-b^2}X_2 B_2^\pm\right|^2\times\\
\delta(E\!-\!E_1\!-\!E_2\!-\!E_+)
\delta(p_{y}\!-\!p_{1y}\!-\!p_{2y}\!-\!p_{+y})\times\\
\delta(p_{z}-p_{1z}-p_{2z}-p_{+z})
d^2p_1d^2p_2d^2p_+,
\end{array}
\end{equation}
where
\begin{equation}
\label{M}
M=\frac{\alpha^2(\tilde m -\mu m)}
{2^7\pi^3 m^2 \tilde m EE_1E_2E_+ hl!}.
\end{equation}

The integration over $d^2p_+$ can be easily performed
using the $\delta$-functions
$\delta(p_y-p_{1y}-p_{2y}-p_{+y})\delta(p_z-p_{1z}-p_{2z}-p_{+z})$.
The probability then takes the form
\begin{equation}
\label{dW_Y}
\begin{array}{l}
dW=2m^2hM\left[\left((B_1^\pm)^2+(B_2^\pm)^2\right)Y-2B_1^\pm B_2^\pm Y'\right]
\times\\
\delta(E-E_1-E_2-E_+)dp_{1z}dp_{2z},
\end{array}
\end{equation}
where we introduced the designations
\begin{equation}
\label{Y13}
\begin{array}{l}
Y=\int\int da\,db\left|e^{-a^2}X_1\right|^2,\\
Y'=\int\int da\,db \, e^{-a^2-b^2}Re(X_1X_2^*).
\end{array}
\end{equation}
The quantity $Y'$ defines the interference term in the
probability of the process.

In Eq.~(\ref{dW_Y}), we will transform the $\delta$ function of the
particle energies to the $\delta$ function of the momentum
components
\begin{equation}
\label{delta}
\delta(E-E_1-E_2-E_+)=\frac{m\sum\limits_{j=\pm}{\delta(p_{1z}-g_j)}}
{\sqrt{4m\Delta-3p_{2z}^2}},
\end{equation}
where
\[g_\pm=\frac{1}{2}\left(-p_{2z}\pm\sqrt{4m\Delta-3p_{2z}^2}\right).\]

In view of the chosen conditions (\ref{h}), (\ref{lf0}), and (\ref{pf}),
the dependence of the probability on the $z$ momentum
components can be neglected everywhere, except the
factors $B_1^-$ and $B_2^-$, and the $\delta$ function (\ref{delta}).
Therefore, the probability (\ref{dW_Y}) can be easily integrated in
finite form. As a result, we will obtain the expressions
\begin{equation}
\label{Wtotal+}
W^+=\frac{\alpha^22m}{3\pi^2\sqrt{3} l!}(Y-Y'),
\end{equation}
\begin{equation}
\label{Wtotal-}
W^-=\frac{\alpha^2 4\Delta}{9\pi^2\sqrt{3} l!}(2Y+Y').
\end{equation}

Let us calculate $Y$ and $Y'$. First of all, note that we
can generally assume from physical considerations
that $a\sim \xi$. Therefore, the middle term with $\xi^{l/2}$
makes a major contribution in the expansion of the binomial
in Eq.~(\ref{intX}). In addition, a numerical analysis of this
expression shows that the principal-value integral can
be neglected compared to the pole residue. Using
these assumptions, we can easily calculate the integral
$X_1$ and then obtain the following result for $Y$ and $Y'$:
\begin{equation}
\label{Y1_res}
\begin{array}{l}
\displaystyle
Y=4\sqrt{2}\pi^2\Omega^{2l}e^{-2\Omega^2}\frac{l!}{l\Gamma(l/2+1)^2},\\
Y' \ll Y,
\end{array}
\end{equation}
where $\Gamma$ is the gamma function.

Substituting $Y$ into Eqs.~(\ref{Wtotal+}) and (\ref{Wtotal-}) and
neglecting the interference term, we will obtain the final
expressions for the total probability of the process
(per unit time, $s^{-1}$)
\begin{equation}
\label{Wp}
W^+=\alpha^2\left(\frac{mc^2}{\hbar}\right)\frac{8\sqrt{2}}{3\sqrt{3}}
\,\, \frac{\Omega^{2l} e^{-2\Omega^2}}{l\Gamma(l/2+1)^2},
\end{equation}
\begin{equation}
\label{Wm}
W^-=\alpha^2\left(\frac{mc^2}{\hbar}\right) \frac{\Delta}{m}
\frac{32\sqrt{2}}{9\sqrt{3}}
\,\, \frac{\Omega^{2l} e^{-2\Omega^2}}{l\Gamma(l/2+1)^2}.
\end{equation}

\section{ANALYSIS OF THE PROBABILITY}
Let us analyze the result obtained. First of all, note
that the total probability contains no singularities,
when the longitudinal particle momenta are zero,
characteristic of the photoproduction process
$\gamma \rightarrow ee^+$ \cite{Shabad, HardingRep, Novak}.

From Eqs.~(\ref{Wp}) and (\ref{Wm}), it is easy to derive the
ratio of the probabilities
\begin{equation}
\frac{W^-}{W^+}=\frac{4}{3}\frac{\Delta}{m},
\end{equation}
where $\Delta=E-3m$. As was pointed out previously,
$\Delta \lesssim mh$ near the process threshold and, hence,
$W^- \ll W^+$. In the special case where the magnetic field is
$h=4/l$, the equality $\Delta=0$ holds and, hence, $W^-=0$ (within
the accuracy of the approximation).

\begin{figure}[h]
\resizebox{\columnwidth}{!}
{  \includegraphics{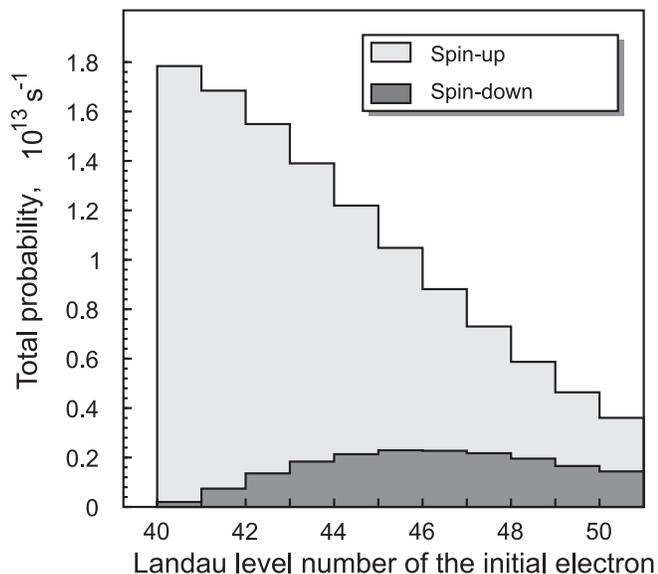} }
\caption{Total probability of pair production at the ground
levels versus Landau level number for the initial electron,
$h=0.1$.}
\label{Total}
\end{figure}

In Fig.~\ref{Total}, the total probability is plotted against the
Landau level number for the initial electron. The magnetic field
is taken to be $h=0.1$, with the threshold value of the Landau
level for the initial electron being $l=40$.
As we see, the probability of the process is $10^{13}$~$c^{-1}$:
\begin{equation}
W^+\sim 10^{13}\,\, c^{-1}.
\end{equation}
Both probabilities decrease with increasing number $l$
and $W^-$ approaches zero near the threshold.

In conclusion, let us compare the probability of the
process considered with the probabilities of other processes.
The table gives the probabilities of the following processes:
emission,
photoproduction,
double synchrotron emission (this process in the field of a
laser wave was considered in \cite{Voroshilo,Lotstedt}),
photoproduction with photon emission,
and pair production by an electron.
The magnetic field is $h=0.1$ ($\approx 4.4\cdot 10^{12}$~G).

\begin{table*}
\caption{\label{tab} Comparison of the probabilities
of various processes, $h=0.1$.}
\begin{ruledtabular}
\begin{tabular}{p{1.5cm}p{1.5cm}p{2.4cm}p{2cm}p{2cm}p{2.5cm}}
 &a &b &c &d &e \\
\hline
Process&
Emission&
Photoproduction&
Double emission&
Photoproduction with emission&
Pair production by electron\\
\hline
Diagram &
\resizebox{0.2\columnwidth}{!}{\includegraphics{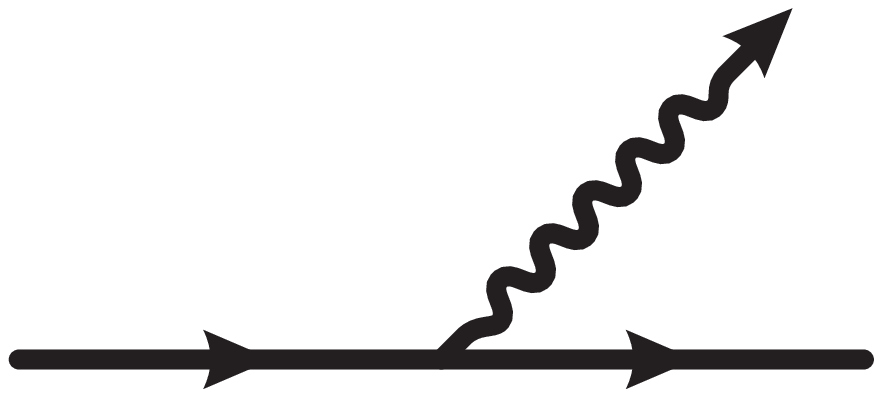} } &
\resizebox{0.2\columnwidth}{!}{\includegraphics{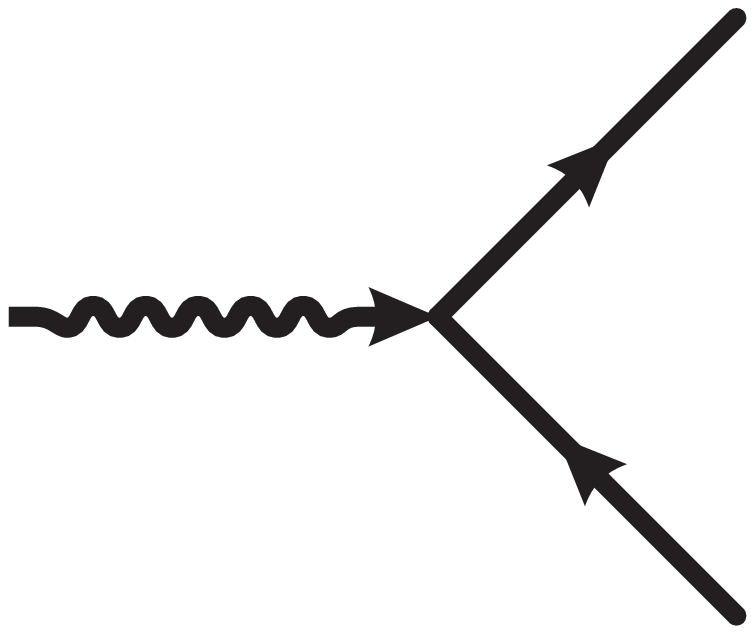} } &
\resizebox{0.2\columnwidth}{!}{\includegraphics{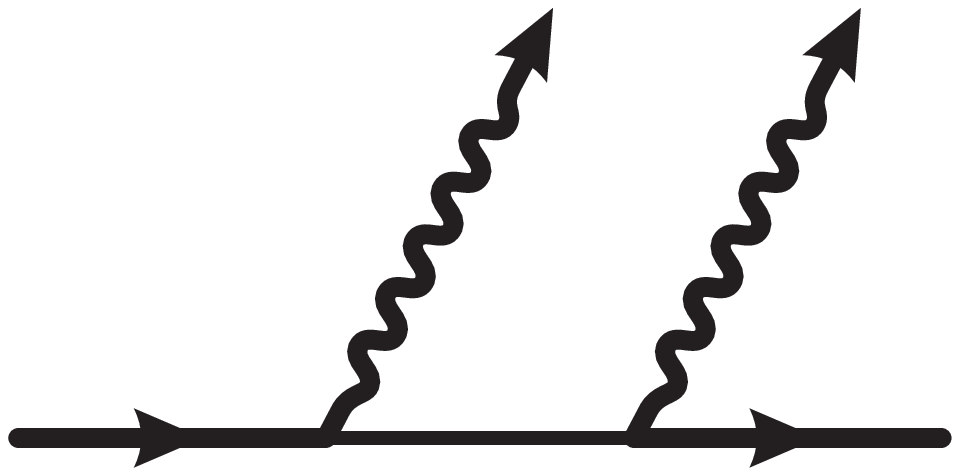} } &
\resizebox{0.2\columnwidth}{!}{\includegraphics{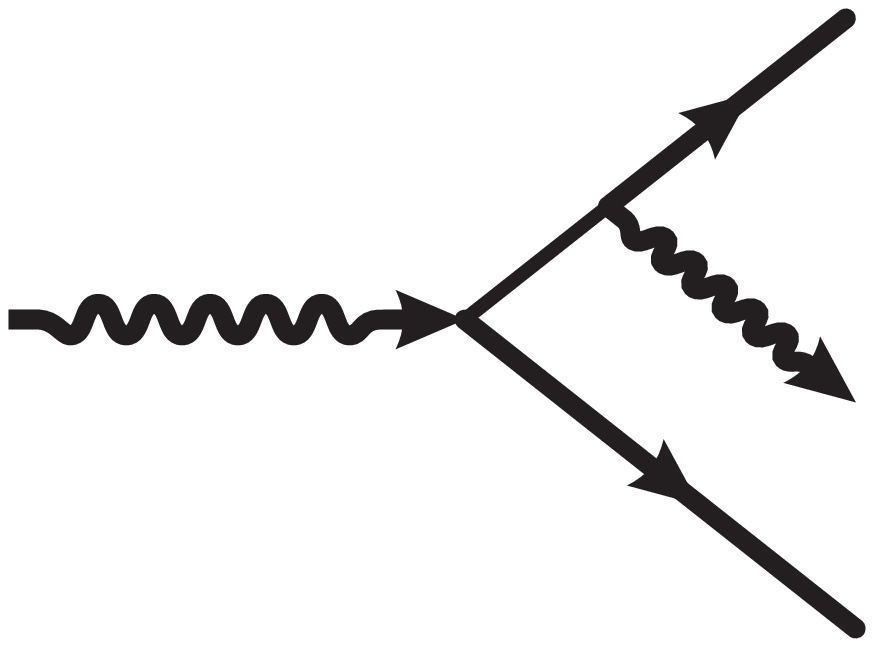} } &
\resizebox{0.2\columnwidth}{!}{\includegraphics{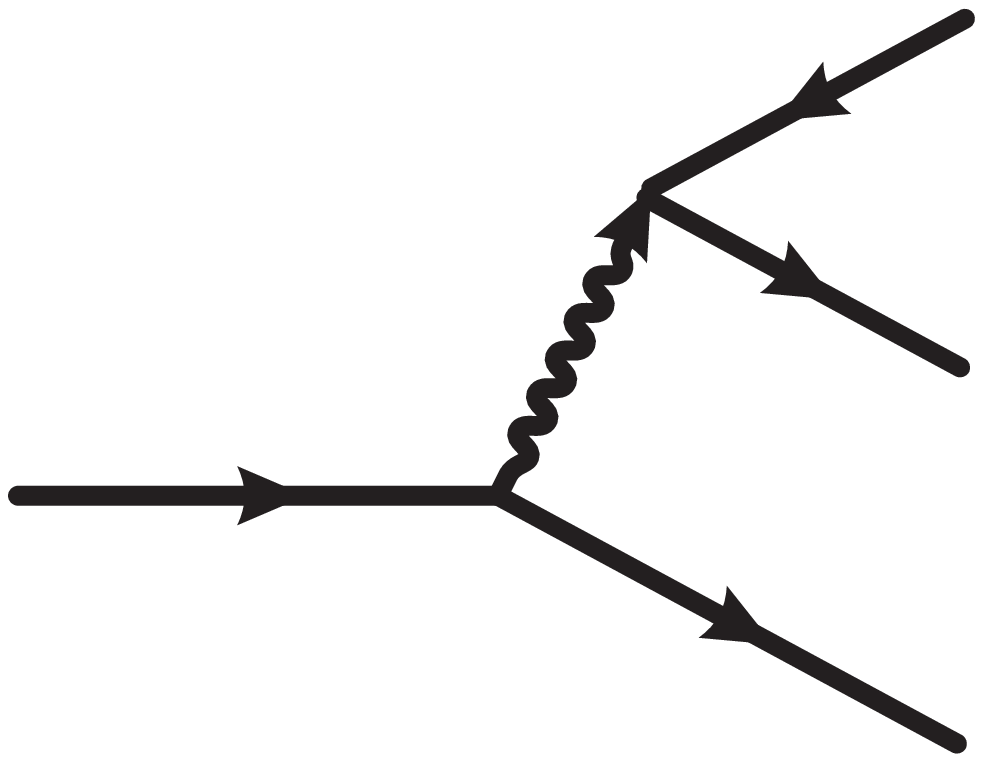} } \\
\hline
Initial conditions &
${l=40}$, ${l'=0}$ &
${\omega=2m}$, ${l_-=l_+=0}$ &
Lowest levels &
Lowest levels &
${l=40}$, ${E=2m}$, ${l_f=0}$ \\
\hline
Probability,~${s^{-1}}$
&
$\begin{array}{l}
\!\!W_{e\rightarrow \gamma e}^{total}\sim 10^{17} \\
\!\!W_{e\rightarrow \gamma e}^{\omega>2m}\sim 10^{14}\\
\!\!W_{e\rightarrow \gamma e}^{1\rightarrow 0}\sim 10^{16}
\end{array}$
&
${W_{\gamma\rightarrow ee^+}\sim 10^{9}}$
&
${W_{e\rightarrow e\gamma\gamma}^{res}\sim
W_{e\rightarrow \gamma e}^{1\rightarrow 0}}$
&
$\begin{array}{l}
\!\!W_{\gamma\rightarrow \gamma ee^+}^{nonres} \sim 10^{6}
\\
\!\!W_{\gamma\rightarrow \gamma ee^+}^{res}\sim
W_{\gamma\rightarrow ee^+}
\end{array}$
&
$W\sim 10^{13}$\\
\hline
References &
\cite{Klepikov, Novak, Kholodov2001}&
\cite{Klepikov,HardingRep,Baier,Novak2008,Novak}&
\cite{Fomin2003}& \cite{Fomin2007}&--\\
\end{tabular}
\end{ruledtabular}
\end{table*}

For the photoproduction probability, we use an
expression derived in the ultraquantum approximation
\cite{Novak2008, Novak}. Let us take the initial photon frequency to be
$\omega=2m$, the electron and positron level numbers to be
$l_f=0$, and the magnetic field to be $h=0.1$. The photoproduction
probability has a resonant pattern and depends significantly on the
$z$ component of the particle momenta. We will choose them to be of the order
of $m\sqrt{h}$ based on our estimate of (\ref{pf}). Then,
$$  W_{\gamma \rightarrow ee^+} \approx 3.7 \cdot 10^{9} c^{-1}.  $$

To estimate the emission probability, it is necessary
to use the ultrarelativistic approximation \cite{Klepikov, Novak}.
Choosing the initial electron energy to be $3m$, we will
obtain an estimate of the total synchrotron emission
probability:
$$  W_{e\rightarrow \gamma e}^{total}\sim 2.8 \cdot 10^{17} c^{-1}. $$
However, this includes the processes of photon
emission with an energy insufficient for pair production.
The probability of emitting a photon with an
energy from $2m$ to $\omega_{max}\approx3m$ is
$$ W_{e\rightarrow \gamma e}^{\omega>2m}\sim6.8 \cdot 10^{14} c^{-1}. $$

\begin{acknowledgments}
We are grateful to S.~P.~Roshchupkin and V.~E.~Storizhko for valuable
remarks and discussions.
\end{acknowledgments}


\end{document}